\documentclass{svproc}
\usepackage[utf8]{inputenc}

\usepackage{url}

\usepackage[sectionbib]{natbib}
\bibpunct{(}{)}{;}{a}{}{,}
\usepackage{graphicx}
\usepackage{amsmath,amssymb}
\usepackage{multirow}
\usepackage{longtable}

\usepackage[b5paper]{geometry}
\def\la{\;
\raise0.3ex\hbox{$<$\kern-0.75em\raise-1.1ex\hbox{$\sim$}}\; }
\def\ga{\;
\raise0.3ex\hbox{$>$\kern-0.75em\raise-1.1ex\hbox{$\sim$}}\; }

\newcommand{\daa}{$\Delta\alpha/\alpha$}
\newcommand{\dmm}{$\Delta\mu/\mu$}
\newcommand{\CI}{[C~{\sc i}]}
\newcommand{\CII}{[C~{\sc ii}]}

\begin{document}
	\mainmatter       
	\title{Fundamental physical constants at low and high redshifts} 
	\titlerunning{Physical constants at low and high redshifts}  
	\author{Sergei A. Levshakov}
	\authorrunning{S. A. Levshakov} 
	\tocauthor{Sergei A. Levshakov}
	\institute{A. F. Ioffe Physical-Technical Institute, St. Petersburg, Russia, \\
		\email{lev@astro.ioffe.ru} }
	\maketitle 

\begin{abstract}
Spacetime variations of physical constants can be associated with the existence
of Higgs-like scalar field(s) that couple non-universally to the baryonic matter.
Recent results of astronomical spectral measurements of the fractional changes in
the electron-to-proton mass ratio, $\mu = m_{\rm e}/m_{\rm p}$, at low ($z \sim 0$)
and high ($z \sim 6.5$) redshifts are discussed. 
It is shown that the distribution of the most accurate
estimates of \dmm~= $(\mu_{\rm obs} - \mu_{\rm lab})/\mu_{\rm lab}$ ranging between
$z = 0$ and $z \sim 1100$ can be approximated by a power low 
\dmm~= $k_\mu (1+z)^p$, with $k_\mu = (1.7 \pm 0.3)\times10^{-8}$ and 
$p = 1.99 \pm 0.03$, implying a dynamical nature of the scalar field(s).
\end{abstract}

\section{Introduction}

The suggestion that the physical constants may vary on the cosmological time
scale can be traced as far back as 1937, when Milne \cite{mil} and Dirac \cite{dir1}
argued about possible changes in the gravitational constant $G$ during the
lifetime of the universe. 
Modern physics considers the masses of quarks and leptons as a result of their
Yukawa coupling to the scalar Higgs field.
Other scalar fields, which are proposed to explain the phenomena of dark energy and
dark matter, can also couple to the standard baryonic matter and, hence, change the masses of
elementary particles.
Since the proton and neutron masses are mainly determined by the binding energy of
quarks \cite{uz}, 
the changes of their masses due to dynamical scalar fields are much smaller
compared to the effect of altering $m_{\rm e}$ \cite{ku:sch}.
Thus, values of the fine structure constant, $\alpha = e^2/\hbar c$, 
and the electron to proton mass ratio, $\mu = m_{\rm e}/m_{\rm p}$,
measured at different physical and cosmological conditions can be used to probe
the hidden scalar fields.
Nowadays, the search for hypothetical variations of $\alpha$ and $\mu$
is among the most extensively studying problems in laboratory and cosmic physics.

The unprecedentedly sensitive limits on the temporal drift of
these dimensionless constants achieved for the passed decades
in laboratory experiments are of one part in
$10^{15}-10^{16}$ per year \cite{bla,fer,sch}.

Astrophysical studies of extragalactic objects at low ($z \sim 0$) and high ($z \sim 1-7$)
redshifts constrain the fractional changes in
$\Delta \alpha/\alpha = (\alpha_{\rm obs} - \alpha_{\rm lab})/\alpha_{\rm lab}$ and 
$\Delta \mu/\mu = (\mu_{\rm obs} - \mu_{\rm lab})/\mu_{\rm lab}$ 
at the same order of magnitude \cite{mol,aga,kan,lev19} 
if we assume a linear drift of the physical constants with cosmic time.
New optical spectral observations with two high-resolution spectrographs
ESPRESSO/VLT \cite{lee} and HIRES/E-ELT \cite{mar} are planning to
put ever deeper constraints on $\alpha$- and $\mu$-variations. 

A recent analysis of the ionization processes at redshift $z \sim 1100$
which are responsible for the temperature and polarization anisotropies of the
cosmic microwave background (CMB) radiation has shown that
the value of the Hubble constant $H_0$ correlates with the electron mass \cite{h:c}.
As a result, an increased effective electron mass at the epoch of recombination,
$m_{{\rm e},z} = (1.0190\pm0.0055)m_{\rm e,0}$, yields
$H_0 \simeq 71$ km~s$^{-1}$~Mpc$^{-1}$~--- the value shifted with respect to
$H_0 = 67.4\pm0.5$ km~s$^{-1}$~Mpc$^{-1}$ which is based on the 
standard $\Lambda$CDM model \cite{planck}.
Such a 2\% increasing of the electron mass at $z \sim 1100$ may alleviate
the difference between the Hubble constant measured at
$z \la 1$, $H_0 = 74.03\pm1.42$ km~s$^{-1}$~Mpc$^{-1}$ \cite{ries}, 
and the CMB value.

\section{Constraints on the time variation of $\mu$}

Note that the spacetime variation of the electromagnetic fine structure constant $\alpha$
should be much smaller than that of $\mu$ \cite{yoo}.
In the following we will consider $\alpha$ unchangeable and put \daa~$= 0$.

The variation of $\mu$ can be probed by comparing the 
relative frequencies of different molecular transitions 
as was originally suggested in 1975 by Thompson \cite{thom}. 
Later on, it was shown that transitions within the molecular bands
have different sensitivities to the variation of $\mu$ \cite{var:lev}. 
The corresponding sensitivity coefficients $Q$ for the Lyman and Werner transitions of H$_2$ 
are typically $\sim 0.01$ and differ in sign \cite{var:lev,ub1}. 
The H$_2$ absorption lines detected in quasar spectra \cite{lev:var,bal} give the following estimation of \dmm:
\begin{equation}
\Delta\mu/\mu = (R_{ij} - R^0_{ij})/[R^0_{ij}(Q_j - Q_i)]\ ,
\label{Eq1}
\end{equation}
where $R_{ij} = (\lambda_i/\lambda_j)_z$ is the ratio of two H$_2$ lines at the epoch $z$, $R^0_{ij}$~--
at $z = 0$, and $Q_i, Q_j$ are the sensitivities of the wavelengths $\lambda_i, \lambda_j$ to the
variation of $\mu$.  
This so-called $H_2\, method$, being applied to optical spectra of quasars, 
provides the mean value of $\langle \Delta\mu/\mu \rangle = (-2.5\pm5.3)\times10^{-6}$
in the interval from $z = 2.05$ to $z = 4.22$ \cite{lev20}.

\begin{figure}
\vspace{-1.0cm}
\includegraphics[width=10.5cm]{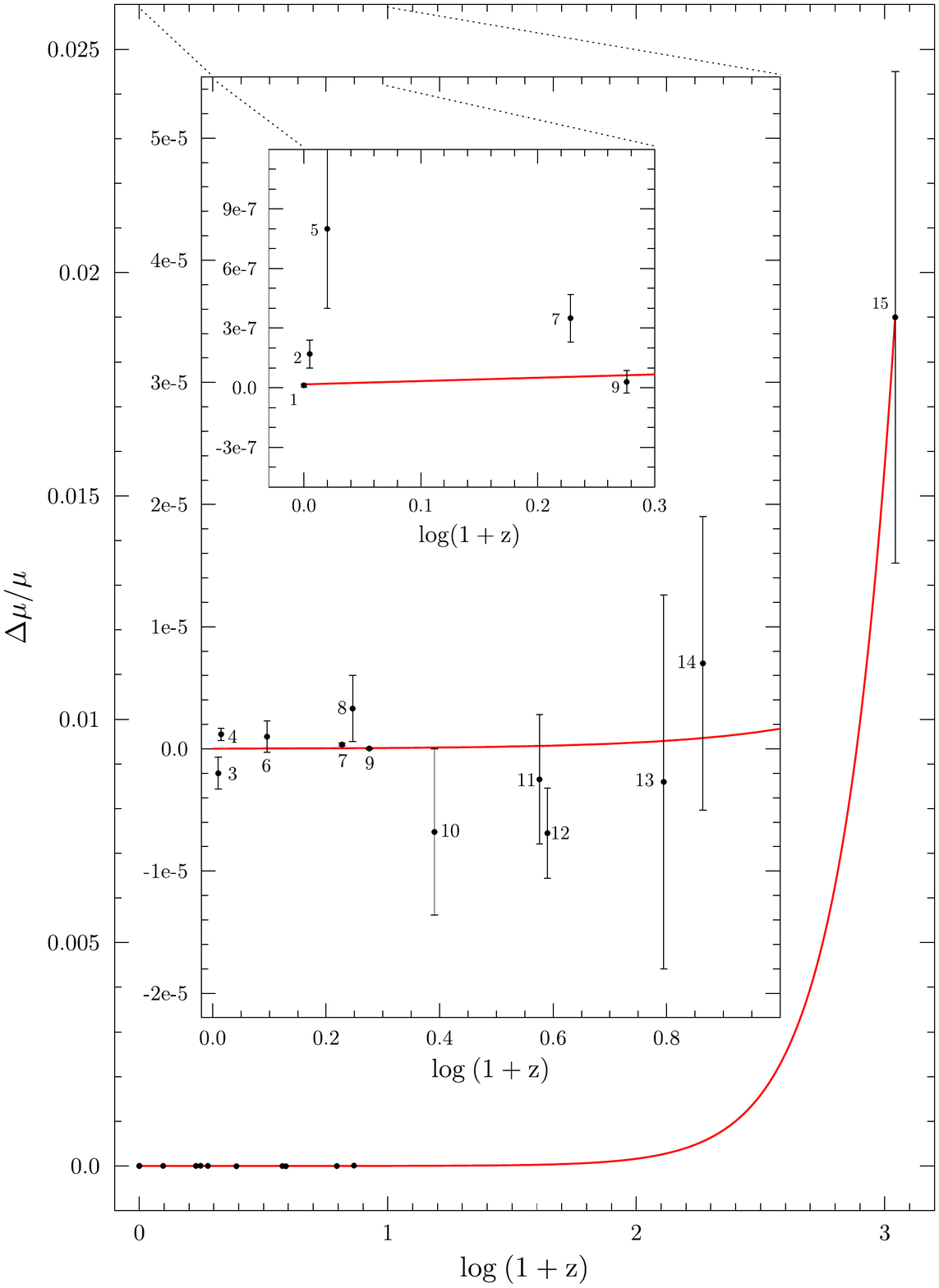}
\vspace{-0.5cm}
\caption{Constraints on the fractional changes in $\mu$ (dots with $1\sigma$ error bards)
as a function of redshift $z$ in units of log$_{10}(1+z)$. 
Two inserts zoom consequently the corresponding parts of the data sample using
different horizontal and vertical scales.
Shown by red is a two-parameter
regression curve \dmm~= $k_\mu (1+z)^p$ with $k_\mu = (1.7 \pm 0.3)\times10^{-8}$
and $p = 1.99 \pm 0.03$ ($1\sigma$). Data points 1,2,6-15 are taken from \cite{lev20},
and 3-5 from the present paper.
Points 1-5 at $z = 0$ are slightly shifted with respect to each other to resolve blending.
}
\label{F1}
\end{figure}

The microwave and submillimeter astronomical spectra are essentially more
sensitive to variations in $\mu$ than the H$_2$ lines \cite{koz:lev}.
For instance, the $NH_3\, method$ \cite{fla:koz} yields
\begin{equation}
\Delta\mu/\mu = (V_{\rm rot} - V_{\rm inv})/[c(Q_{\rm inv} - Q_{\rm rot})]\ , 
\label{Eq2}
\end{equation}
where the observed radial velocity $V_{\rm inv}$
of the inversion transition of NH$_3$(1,1) 
with $Q_{\rm inv} = 4.46$
is compared with a suitable rotational radial velocity $V_{\rm rot}$ of another molecule
co-spatially distributed with ammonia and having $Q_{\rm rot} = 1$, and $c$
is the speed of light.
In the Milky Way disk, the $NH_3\, method$ provides
 $\langle \Delta\mu/\mu \rangle = (2\pm6)\times10^{-9}$ \cite{lev13}.

A wide range of $Q$-values (from $\simeq -15$ to $\simeq 45$) of methanol CH$_3$OH lines
was deduced in \cite{jan,lev11}. 
Any two transitions of CH$_3$OH 
with different sensitivity coefficients $Q_i$ and $Q_j$
give the following estimate of \dmm: 
\begin{equation}
\Delta\mu/\mu = (V_j - V_i)/[c(Q_i - Q_j)]\ , 
\label{Eq3}
\end{equation}
where $V_j$ and $V_i$ are the apparent radial velocities of the
corresponding methanol transitions. 
The deepest limits on \dmm\ were obtained by the
$CH_3OH\ method$ for the Galactic cloud L1498 \cite{dap} and the $z = 0.89$ gravitational
lens \cite{kan}: \dmm~= $(-3\pm2)\times10^{-8}$ and $(3\pm6)\times10^{-8}$, respectively.
 
Finally, to probe \dmm\ by spectroscopic methods at very high redshifts ($z \ga 6$)
a fine structure transition method (FST) was proposed in \cite{lev08}.
It is based on the analysis of the radial velocity offsets between CO rotational lines
and \CI\ and/or \CII\ fine-structure transitions. Under assumption that  \daa~$= 0$,
it gives \cite{lev20}:
\begin{equation}
\Delta\mu/\mu = \Delta V/c\ , 
\label{Eq4}
\end{equation}
where $\Delta V$ is the radial velocity offset, $\Delta V = V_{\rm fs} - V_{\rm rot}$. 
Three estimates of \dmm\ at $z = 6.003$, 6.419, and 6.519 by the FST method provide a weighted
mean value of 
$\langle \Delta\mu/\mu \rangle = (0.7\pm1.2)\times10^{-5}$
at $\bar{z} = 6.3$ \cite{lev20}.

All available values of \dmm, ranging between $z = 0$ and $z \sim 1100$, were put together in Fig.~3 in \cite{lev20}.
Here, in Fig.~\ref{F1}, we reproduce this compilation with a
slightly updated dataset: instead of one \dmm\ value averaged over the whole Triangulum galaxy M33 ($D \simeq 800$ kpc)
we use three \dmm\ values calculated on base of the CO(2-1) and \CII\ emission lines in regions 
at different galactocentric distances: $R_c = 0, 2$, and 3.5 kpc.
The distribution of the updated data points 
(shown by dots with $1\sigma$ error bars in Fig.~\ref{F1}) was approximated by a simple power law
\begin{equation}
\Delta\mu/\mu = k_\mu (1 + z)^p\ ,
\label{Eq5}
\end{equation}
which gives $k_\mu = (1.7 \pm 0.3)\times10^{-8}$ and $p = 1.99 \pm 0.03$ ($1\sigma$). 
The corresponding regression curve is shown by red in Fig.~\ref{F1}.

An important point to note is that 
the revealed $z$-dependence of $\mu$ crucially depends on the CMB estimate of the electron mass
at $z \sim 1100$.

It is obvious that to verify the redshift dependence of $\mu$ a more refined
analysis of the CMB anisotropies is needed along with more accurate spectral measurements at lower redshifts.
The required uncertainty of the molecular line position measurements should be less or about 10 m~s$^{-1}$
which can be achieved with the existing radio astronomical facilities in observations of objects
in the local Universe ($z \la 1$).


\bibliographystyle{aa}
\bibliography{template}
\end{document}